\documentclass[3p,times]{elsarticle}

\usepackage{graphicx}
\usepackage{xcolor}

\journal{Solid State Communications}

\usepackage{amssymb}

\begin{document}

\begin{frontmatter}




\title{Bi, Cr and Ag dopants in PbTe and SnTe: impact of the host band
  symmetry on doping properties by ab initio calculations}


\author[a]{A. {\L}usakowski}
\author[a]{P. Bogus{\l}awski}
\author[a,b]{T. Story}
\address[a]{Institute of Physics, Polish Academy of Sciences,
  Al. Lotnik\'ow 32/46, PL-02-668 Warszawa, Poland}
\address[b]{International Research Centre MagTop, Institute of 
Physics, PAS, Al. Lotnik\'ow 32/46, PL-02-668,
 Warsaw, Poland}

\begin{abstract}


Doping properties of Bi, Cr and Ag dopants in thermoelectric and  
topological materials PbTe and SnTe are analyzed based on density 
functional theory  calculations in the local density  approximations and the 
large supercell  method. In agreement with experiment, in both PbTe  and 
SnTe, Bi is a donor and Ag is an acceptor with a vanishing magnetic  
moment. In  contrast, Cr is a resonant donor in PbTe, and an resonant 
acceptor in SnTe. We also consider the electronic structure of cation 
vacancies in PbTe and SnTe, since these abundant native defects induce 
$p$-type conductivity in both hosts. 
The quantitatively different impact of these dopants/defects on the host 
band structure of PbTe and SnTe (level energies, band splittings, 
band inversion, and a 
different level of  hybridization  between dopant and host states) is 
explained based on the  group-theoretical arguments.

\end{abstract}

\begin{keyword}
  IV-VI semiconductors $\sep$ band structure \sep dopants



\end{keyword}

\end{frontmatter}
\ 



\section{Introduction}
\label{intro}

PbTe, SnTe, and their substitutional alloys (Pb,Sn)Te are IV-VI 
semiconductor materials known for their intriguing physical properties 
(topological crystalline insulators, ferroelectrics, nonparabolic materials) 
and applications in thermoelectricity and infrared optoelectronics 
\cite{nimtz, khoklov, hsieh, dziawa, tanaka, xu, dybko, osinniy}. For both 
research and application purposes, the vital element of success is a reliable 
$n$- and $p$-type doping extending over several orders of magnitude: while 
optimal thermoelectric properties are observed for heavy bulk doping $n,p$ 
$\sim 10^{19}-10^{20}$~cm$^{-3}$, for topological research we strongly 
prefer very low carrier concentration $n,p$ $\sim 10^{16}- 10^{17}$~cm$^{-
3}$ with negligible contribution of bulk conductivity. 

Doping properties of a given atomic species strongly depend on the host 
band structure and its chemical composition. In most III-V and II-VI 
semiconductors, the conduction band minimum (CBM) has the $\Gamma_1$ 
symmetry, while the valence band maximum (VBM) the $\Gamma_{15}$ 
symmetry (when spin is neglected). In consequence, electronic structure of 
donors and acceptors is quite different, reflecting different degeneracies, 
wave functions, and effective masses of the corresponding free carriers. 
These "classical" issues are extensively discussed in e.g. \cite{yu_cardona}. 
Finally, different impurity energies relative to the VBM in various hosts are 
to some extent governed by the valence band offset \cite{Zunger}. 

While the search for efficient dopants of IV-VI compounds was successful, a 
deeper understanding of the physics of doping is lacking. In these 
semiconductors, the VBM and CBM are at the L-point 
of the Brillouin zone (BZ), and both 
the valence band and the conduction band have similar features: they are 
orbital singlets with ellipsoidal valleys and similar effective masses. 
However, as we show, the fact that they are of opposite parity with respect 
to inversion has a decisive impact on the way they respond to dopants. 
Interestingly, this feature holds also in the case of native defects -- the 
cation vacancies. 

For our analysis we have chosen two good thermoelectric materials: SnTe 
(a topological crystalline insulator) and PbTe (a topologically trivial material 
with very strong spin-orbital interactions).
PbTe-SnTe materials are known to be efficiently doped with traditional
donor (I) or acceptor (Na or Cl) \cite{hoang}. These simple
dopants allow for achieving thermoelectrically optimal carrier
concentration, however, they do not provide new opportunities to
further improve thermoelectric performance, e.g. by electron or phonon
structure engineering. A special class of resonant dopants constitute
group III (Al, Ga, In, Tl) elements \cite{ahmad, Tan, Jovovic,
  Xiong} Our choice of dopants analyzed in this work 
aims at materials less studied, but already of practical importance,
that constitute physically intriguing systems, as explained
below. \\
Ag is a constituent material of PbTe -- AgSbTe$_2$ alloys (known as LAST
thermoelectrics) \cite{snyder} with top heat-to-electricity conversion efficiency
used in thermoelectric generators operating in mid-temperature
range. Truly widespread use of these acceptors requires, however,
thorough understanding of such issues as mutual solubility of lead,
tin, and silver  tellurides, electrical activity of  Ag in
multicomponent alloys with Pb and Sn, as well as the effect of very
heavy doping with Ag on  electronic states in the band gap region. 
\\
Bi is known as an efficient donor center in PbTe bulk crystals and
epitaxial layers, thus permitting reaching electron concentrations
exceeding $n$=10$^{20}$ cm$^{-3}$ . In contrast, much less is known about Bi
electrical activity in SnTe (no $n$-type doping was realized, 
possibly because of its competition
with abundant native defects -- Sn vacancies). 
A new theoretical insight
is needed fully taking into account very strong relativistic effects
(like spin-orbit interactions), which magnitude in Bi (the heaviest
non-radioactive element) could be  higher  than the fundamental band
gap of narrow-gap IV-VI semiconductors, therefore influencing not only
the band gap $E_{gap}$, but also the symmetry of conduction and
valence bands.  
\\
Although doping of IV-VI semiconductors with the 3d transition metal Cr
was experimentally studied in PbTe:Cr
\cite{story92,story93,grodzicka,paul,nielsen, wang2} and PbSe:Cr
already three 
decades ago, a complete theoretical description is still lacking,
especially in SnTe:Cr and Pb$_{1-x}$Sn$_x$Te:Cr  crystals with the inverted band
ordering. The key physical challenge concerns proper  accounting for
an unusual energy location of Cr-derived 3d orbitals being in
resonance with either conduction band (in PbTe:Cr, well established
experimentally) or with valence band (expected in
SnTe:Cr). Particularly important is the interaction/hybridization
between $3d$(Cr) orbital states and valence or conduction band states of
a given symmetry (changing upon band inversion in  Pb$_{1-x}$Sn$_x$Te:Cr with
a high enough Sn content \cite{skipetrov}).

In the literature there are many papers devoted to the dopants in
IV-VI semiconductors (see for example review \cite{wiendlocha}).
In general, in most of the papers, the analysis  is
devoted to changes in the 
density of states (DOS) caused by impurities
  \cite{ahmad,mahanti,ahmad1} . The additional peaks of DOS may 
significantly influence the thermoelectric power factor $P_{\sigma}=\sigma S^2$ what leads
to improved thermoelectric figure of merit
$ZT=P_{\sigma}T/\kappa$ where $S$, $\sigma$, $\kappa$ and $T$
are Seebeck coefficient, electrical conductivity, thermal conductivity and
temperature, respectively.
\\

After introducing in Section \ref{methods} the methods of calculations and 
some technical details, we begin Section \ref{res} with the analyses of band 
structures of pure PbTe and SnTe as a starting point, calculated 
using the large supercells to show the unperturbed energy bands. We then 
move to the cation vacancies in PbTe and SnTe. They are the dominant 
native defects in both PbTe and SnTe \cite{nimtz,khoklov}, and they 
determine the $p$-type conductivity in both intrinsic hosts. Next, we study Pb 
ion substituting Sn in SnTe -- a simple case of an isoelectronic impurity, 
which induces a weak perturbation and only small modifications of the 
energy bands. This case is considered as a reference, because it helps 
understanding not only formation of energy bands in the Pb$_{1-x}$Sn$_x$Te alloy, 
but also the effects present in the remaining, more complex, cases. In 
contrast, substitutional Bi induces an impurity band in the band gap and 
acts as a donor. Finally, the transition metal impurities, Cr and Ag, are 
considered. Section \ref{summary} summarizes the main results.

\section{Methods} 
\label{methods} 
\begin{figure}
 \includegraphics[width=\linewidth]{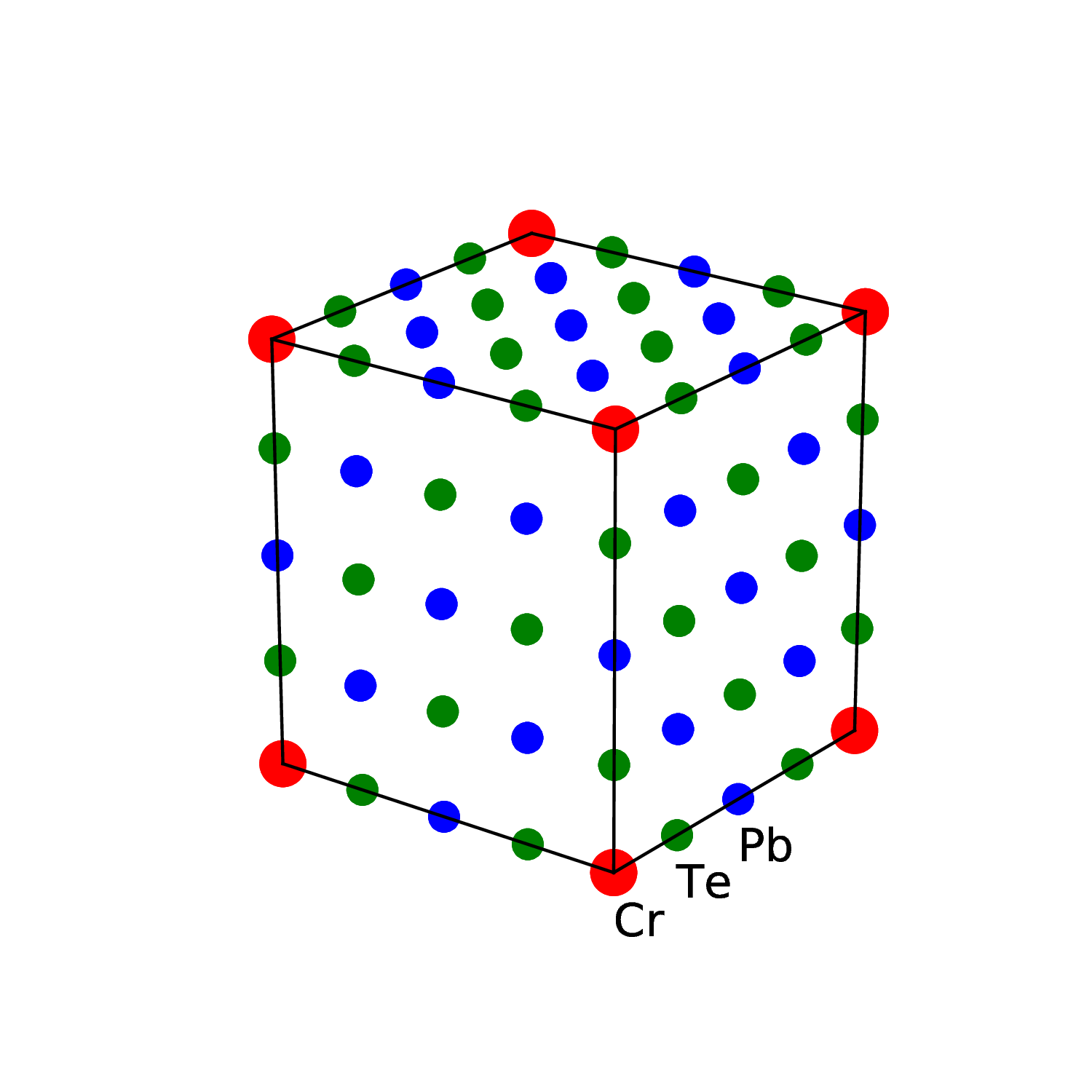}
 \caption{
   A scheme of a $2\times2\times2$ supercell with 64 atoms. 
   Importantly, for the sake of clarity only 
   the atoms at the supercell front faces are shown. Impurity atoms are at the cube corners. 
   A larger $3\times 3\times 3$ supercell is used in the calculations.}
 \label{fig1}
\end{figure}

The density functional theory (DFT) calculations are done using the local 
density approximation with the Ceperly-Alder exchange-correlation 
functional \cite{CA}. We employ the OpenMX open-source package 
\cite{openmx}, which also provides the fully relativistic atomic 
pseudopotentials. To obtain proper band
  structure for PbTe we apply the possibility offered by OpenMX of
  controling spin-orbit coupling strength for $6p$ orbitals of Pb. Its 
  change to 0.554 of the value in the Pb atom results in the correct 
  value of the band gap and symmetries of conduction
  and valence bands wavefunctions.  We use the supercell geometry, and consider
$3\times 3\times 3$
cubic supercells containing 216 atoms with periodic conditions imposed in 
all directions. A smaller $2 \times 2 \times 2$ supercell with 64 atoms is shown in
Figure \ref{fig1} as illustration. 
Substituting one cation, Pb or Sn, by a dopant atom corresponds to the 
concentration of impurities about 1 atomic percent of cations, i.e., to the 
concentration of about 10$^{20}$ cm$^{-3}$. One can thus expect that the 
dilute limit is realized, and the results reasonably approximate the features 
of an isolated impurity. The $k$-point sampling for the supercell Brillouin 
zone (BZ) is $2\times 2\times 2$. For both hosts with dopants or with vacancies, the 
optimization of atomic positions is performed. The calculations are stopped 
when the forces acting on atoms are smaller than 5$\times10^{-4}$ 
Ha/bohr, and total energy is converged to within $10^{-6}$ Ha. In the 
calculations we assume the experimental lattice parameters, 6.46 \AA\ for 
PbTe and 6.30 \AA\ for SnTe \cite{nimtz,khoklov}, and we do not take into 
account the small changes of lattice parameters induced by impurities. 
After completing selfconsistent calculations, OpenMX provides a set of 
parameters, which are used in the tight binding approximation calculations 
of the band structure, DOS, and partial density of states 
(PDOS). 
In the supercell approach, the increase in the lattice constant implies that 
the corresponding Brillouin zone is reduced by the same amount. The 
reduction occurs through "folding", which increases degeneracies at high 
symmetry points of the BZ, and superimposes some original directions onto 
just one. In our case, at each "L" point there are states from the
four  L-points of the BZ of the 2-atom host unit cell, and thus bands
at "L" are 8-fold degenerate. Next, the "L" -- $\Gamma$ direction
contains not only the  
fcc $\Lambda$ direction, but also three others folded onto it. The main 
results of our calculation are the energies of the impurity-induced bands 
and their dispersions, while their exact shapes in various directions are less 
relevant. Indeed, energy bands show the exact band energies relative to 
each other. For this reason, and for the sake of transparency, energy bands 
are shown along one direction only, "L"-$\Gamma$. On the other hand, 
complementary information is provided by DOS and PDOS. They allow for 
establishing the widths of impurity bands, their orbital content, and the 
degree of hybridization with the host states. We note that presenting the 
band structure calculated along "L"-$\Gamma$ direction one covers the most 
important band gap region at the L-point in PbTe, as well as a small shift of 
the positons of the VBM and CBM in SnTe along the L-$\Gamma$ direction.

In the tight binding approximation, the wavefunction of the $n$-th band 
and the wave vector $k$ is
\begin{equation}
\psi_{nk}(r) = \sum_{l,s}a_{nk}^{ls}\chi_{kls}(r),
\end{equation}
where $l$ denotes the orbitals and $s$ the spin (up or down). The function 
$\chi$ is defined as
\begin{equation}
 \chi_{kls}(r) = \frac{1}{\sqrt{N}}\sum_{R,\tau}e^{ikR}\phi_{l}(r-R-\tau_l)|s>, 
\end{equation}
where $R$ numbers the supercells' positions and $\tau_l$ denotes the 
center of $l$-th orbital in the supercell. $N$ is the number of supercells in 
the crystal. 
The contribution of $l_0$-th orbital to the wavefunction $\psi_{nk}$ is 
defined as 
\begin{equation}
 C_{nl_0}(k) = \sum_{l=l_0\in A,s} |a_{nk}^{ls}|^2, 
\end{equation}
where $A$ is either the analyzed dopant or the set of all other host cations 
or anions. In the above sums, neglecting summation over spin, there is only 
one summand if $A$ denotes a dopant, but 107 summands for the host 
cations and 108 for the anions. Below, this orbital's contribution is referred 
to as a projection of the wavefunction onto atomic orbitals.

\section{Results and discussion}
\label{res}

\subsection{Band structure of PbTe and SnTe}
Since the qualitative interpretation of our results is based on symmetry 
arguments, we begin by recalling that in the case of PbTe, the wave functions 
of the VBM are of the $L_6^+$ symmetry, i.e. they are even with respect to 
inversion with the center at the cation site, and are mainly composed of the 
$s$(Pb) and $p$(Te) orbitals. The CBM wave functions are of the $L_6^-$ 
symmetry, odd with respect to inversion, and they are mainly composed of 
$p$(Pb) and $s$(Te) orbitals. In SnTe with the inverted band structure, the order of 
the CBM and VBM (together with their corresponding wave functions) is 
reversed. The calculated band gaps are $E_{gap}$(PbTe)= 0.2 eV and 
$E_{gap}$(SnTe)=0.3 eV, what agrees well with experiment \cite{nimtz, khoklov}. 
The point group symmetry of both PbTe and SnTe is $O_h$, which includes 
inversion. In this case, which also holds for other dopants considered here 
except Cr, the bands are spin degenerate. Accordingly, both energy and 
orbital composition of such spin partners are identical. For this reason, below 
we analyze only wavefunctions of one spin partner. 
\begin{figure}
 \includegraphics[width=\linewidth]{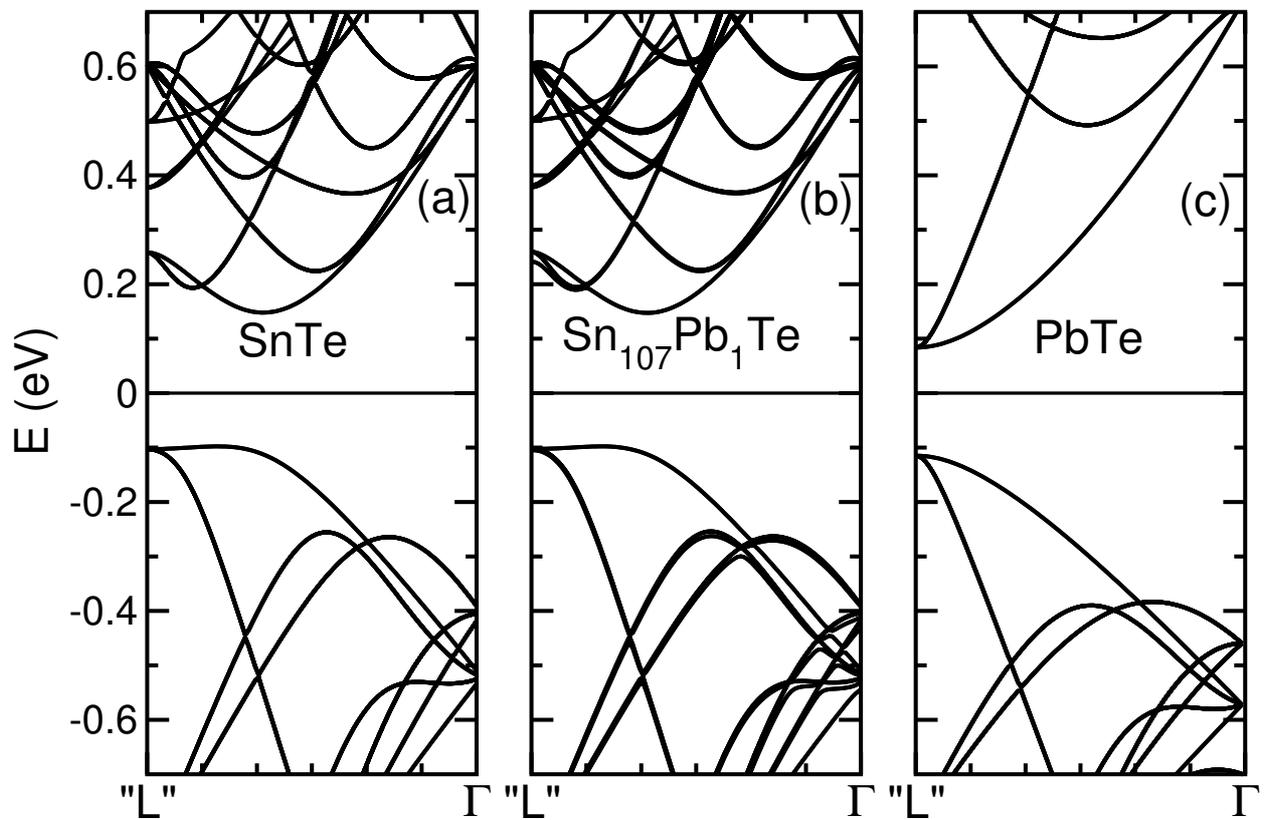}
 \caption{
The calculated energy bands along the "L" -- $\Gamma$ direction in the BZ for 
(a) SnTe, and (c) PbTe. Panel (b) shows the energy bands of SnTe with one Sn 
substituted by Pb, i.e., the Sn$_{107}$Pb$_{1}$Te$_{108}$ supercell. Zero 
energy is at the Fermi level, and is located in the mid-gap position.}
 \label{fig2}
\end{figure}
In Figs \ref{fig2}(a) and \ref{fig2}(c) we show the energy bands for SnTe and PbTe supercells 
along the [111] direction of BZ, respectively. Four 
nonequivalent L-points of the fcc BZ are folded to a point denoted 
by "L" in the supercell BZ. After the folding, both the VBM and 
CBM are 4-fold degenerate at "L" neglecting spin (i.e., 8-fold degenerate with 
spin). One can notice that in SnTe  both CBM and VBM are shifted away
from the L-point, in agreement with the previous findings  
\cite{nimtz, liu}.
The VBM at the L-point is not a local minimum, but a saddle point. In
our work, the band structure of Fig. 2 is plotted along the “L” -- $\Gamma$ 
direction of the BZ of the 216-atom unit cell, therefore actually the
CBM is close to the L-point of the unfolded BZ of the rock-salt
structure.  This distorted E(k) dependence in SnTe is limited to the
VBM region (valence band energy range below 50 meV) and has a minor
influence on hole states at the Fermi level, which in SnTe typically
is in the range 180-250 meV. It is due to  inherently high
concentration of electrically active native defects in SnTe (Sn
vacancies) resulting in conducting hole concentration $p=2\times 10^{20}$ --
$2\times 10^{21}$ cm$^{-3}$. 
\subsection{Cation vacancies in PbTe and SnTe}
\begin{figure}
\includegraphics[width=\linewidth]{figur3.eps}
\caption{
Calculated energy bands of (a) SnTe with a Sn vacancy, and (b) PbTe with a Pb vacancy. 
The dotted line shows the band which is double degenerate and $formally$ 
occupied with one electron. The actual band occupation is determined by the 
Fermi level $E_{Fermi}$, which is taken as zero energy. Here, $E_{Fermi}$ is located below 
the VBM.
}
\label{fig3}
\end{figure}

Figure \ref{fig3} shows the energy bands of SnTe and PbTe supercells with one cation 
vacancy. In either case the host band structure is not strongly perturbed by 
the defect. In both crystals, $V_{cation}$ is a double acceptor 
\cite{nimtz,khoklov}. In PbTe, $E_{Fermi}$ corresponding to zero energy is 
located 0.2 eV below the VBM, and in SnTe it is located at 0.1 eV below the 
VBM. All orbital quartets (neglecting the spin) at the "L" point are split into 
singlets and triplets, but these splittings are of the opposite sign for the 
$L^+_{6}$ and $L^-_{6}$ bands. More specifically, the singlet derived from the 
$L^+_{6}$ state is down-shifted in energy, while that derived from the $L^-
_{6}$ band is up-shifted, whereas the energies of the triplets are almost 
unchanged. As a result, the band gap region at the "L" point of PbTe is 
practically unperturbed by the presence of a vacancy, whereas the gap of 
SnTe is decreased. This can be interpreted tentatively as a formation of the 
acceptor state in the band gap of SnTe. On the other hand, $V_{\mathrm 
{Pb}}$ in PbTe is a resonant acceptor, with a level degenerate with the 
continuum of the valence bands and the wavefunction strongly hybridized 
with the host states. 
\subsection{Isoelectronic Pb$_{\mathrm Sn}$ defect in SnTe}

We now turn to a simple case of a Pb atom substituting Sn in SnTe. The 
Pb$_{\mathrm Sn}$ defect is isolectronic since both Pb and the host Sn 
ions are from the group IV of the Periodic Table. Figure 2(b) displays energy 
bands of SnTe:Pb. One can see that changes in band energies of SnTe are 
about 10 meV. In this case, the perturbation potential (which is the 
difference in the atomic potentials of Pb and Sn in SnTe) is limited to the 
region of the atomic core, and it does not contain the long range Coulomb 
tail. As a result, no gap states are induced. Thanks to the same valency as 
Sn, a Pb substitutional dopant cannot provide free carriers, but it modifies 
the band structure of the SnTe host \cite{nimtz,khoklov}. This case allows 
for an insight into the formation of energy bands of the (Pb,Sn)Te alloy. 
In particular, the impact of 1 at. \% of Pb is quite small, as the energy changes are of a few meV. 
This stems from the similarity between Sn and Pb, which size and electronegativity are close. 
This also provides an argument justifying appropriateness of the virtual crystal 
approximation for the extended band states of (Pb,Sn)Te. In the opposite case of very 
different anions, like O and Te, the virtual crystal approximation leads to fundamental errors.

\begin{figure}
\includegraphics[width=\linewidth]{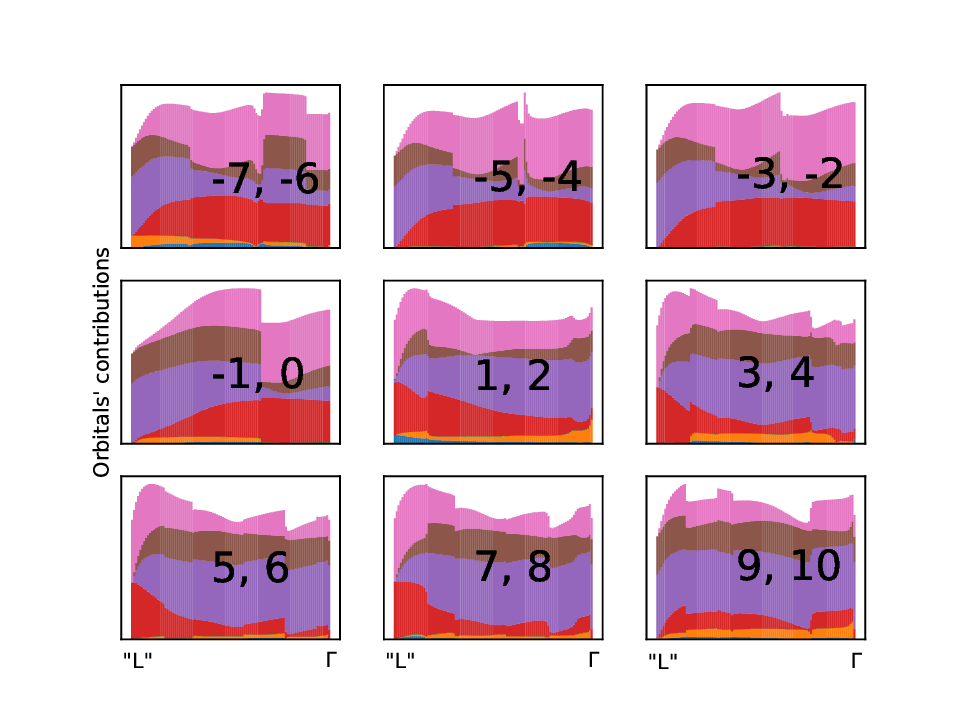}
 \caption{
Projection of the wave functions onto atomic orbitals along the [111] 
direction for SnTe:Pb. The bands from -7 to 0 are valence bands, and from 
1 to 9 are conduction bands. The used color code is as follows: 
blue=6$s$(Pb), orange=6$p$(Pb), green=5$d$(Pb), red=sum of $s$(Sn),
 violet=sum of $p$(Sn), brown= sum of $s$(Te), magenta= sum of
 $p$(Te).
 Because of the inversion symmetry, all bands are double degenerate,
 and the orbital composition of the spin partners (e.g., -7 and -6, -5
 and -4, etc) are identical. 
 }
 \label{fig4}
\end{figure}

As it follows from the Fig. \ref{fig2}(b), the only noticeable impact of 
Pb$_{\mathrm Sn}$ on the bands of SnTe occurs at the CBM in the vicinity 
of the "L" point. The decomposition of the wavefunctions (i.e., projection of the wavefunction onto the 
relevant atomic orbitals) agrees with this result. Indeed, as it follows from Fig. \ref{fig4}, the CBM 
contains a contribution of the 6s(Pb) orbitals, while the 6p(Pb) and 5d(Pb) orbitals do not 
contribute neither to the highest valence bands nor to the lowest conduction bands close to 
"L". 
The discontinuities in the orbital content of most of the wave
functions presented in Fig. \ref{fig4}  occur whenever two bands of
different symmetry intersect  as a function of the wave vector. 
It should be stressed that in Fig. \ref{fig4} we present the sum of 
contributions of, e.g., $5s$(Te) orbitals of $all$ 108 Te ions in the supercell, 
but of only {\em one} $6s$(Pb) orbital, which gives the impression of an 
almost negligible impact of Pb$_{\mathrm Sn}$. This comment holds also 
for other defects discussed below. 

\subsection{Bi donor in PbTe and SnTe} 

\begin{figure}
  \includegraphics[width=\linewidth]{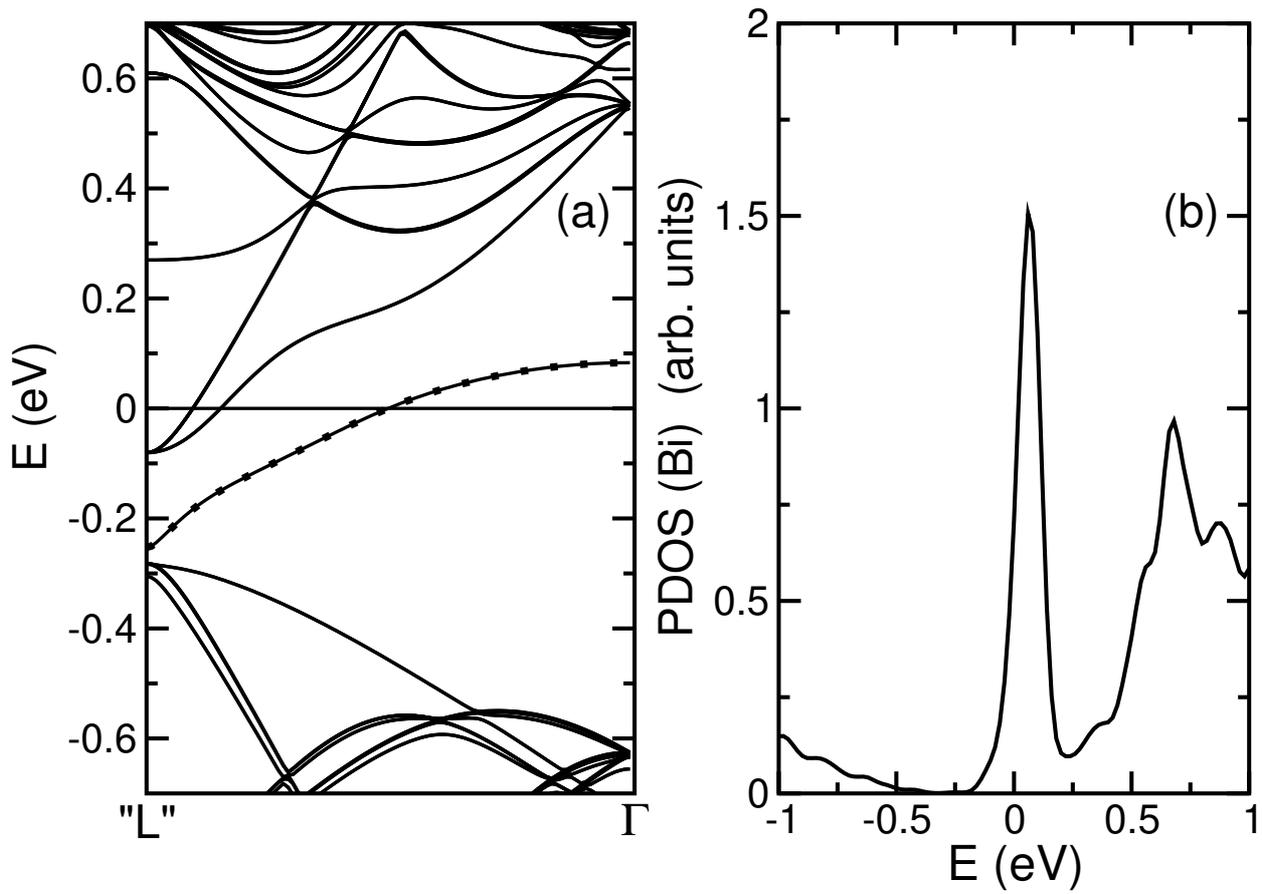}
\caption{(a) Calculated energy bands of PbTe:Bi along the [111] direction. 
The dotted line is double degenerate and $formally$ occupied with one 
electron. The actual band occupation is determined by the Fermi energy. 
Zero energy  is located at the Fermi level.
(b) The corresponding partial density of states of Bi. Zero  energy is at $E_{Fermi}$.}
  \label{fig5}
\end{figure}

\begin{figure}
  \includegraphics[width=\linewidth]{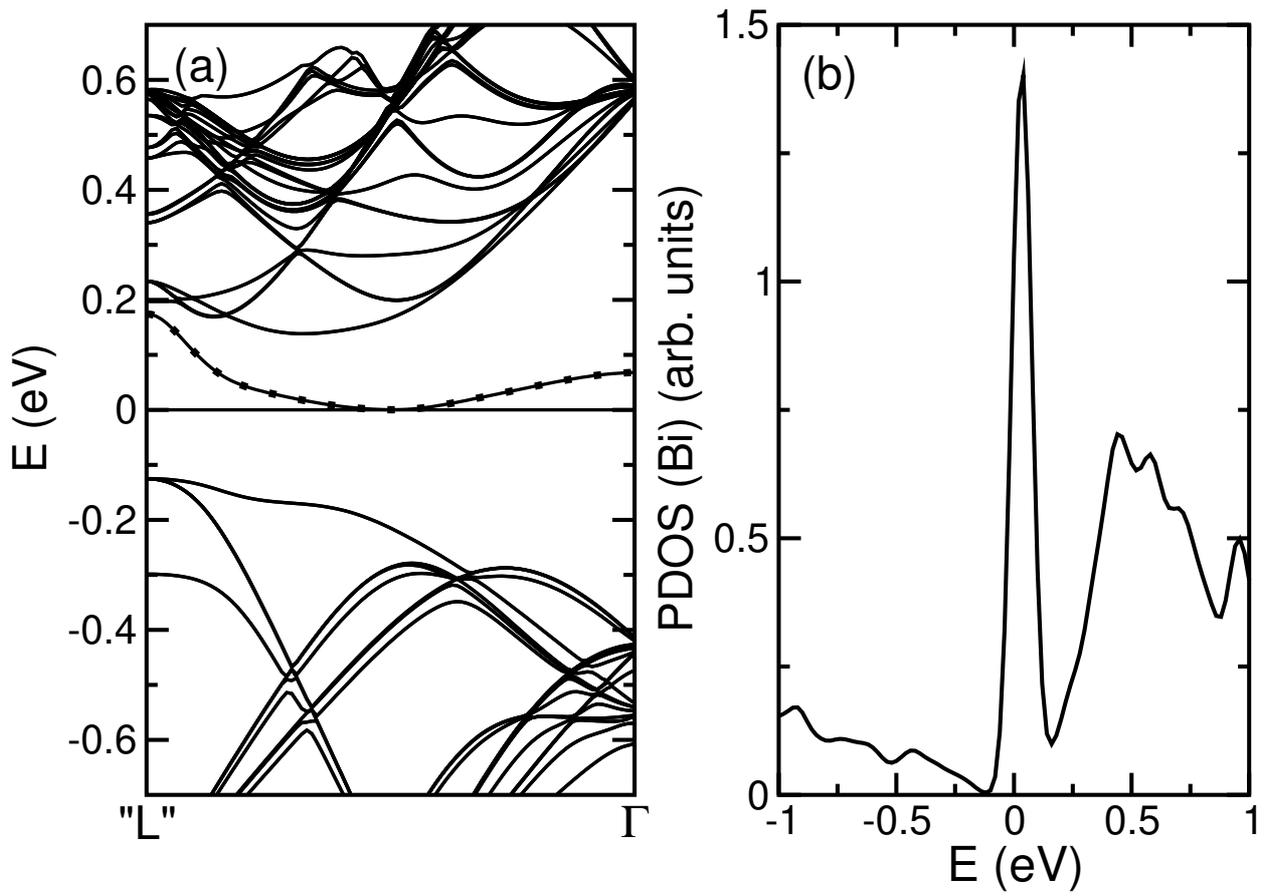}
  \caption{(a) Band structure of SnTe:Bi along the [111]
    direction. The dotted line is double degenerate and $formally$ occupied 
with one electron. The actual band occupation is determined by $E_{Fermi}$. 
(b) Partial density of states of Bi atom. }
  \label{fig6}
\end{figure}

The group-V Bi substituting a host group-IV cation is expected to be a donor.  
The calculated energy bands for both PbTe:Bi and SnTe:Bi are shown in Figs  
\ref{fig5} and \ref{fig6}, respectively. In both hosts, the Bi-derived energy levels 
induce a ~0.3 eV wide impurity band in the band gap, formally occupied with one 
electron, i.e., Bi is a single  donor. Inclusion of atomic relaxations around the 
dopant induces negligible changes in the Bi energy levels.  

\begin{figure}
  \includegraphics[width=\linewidth]{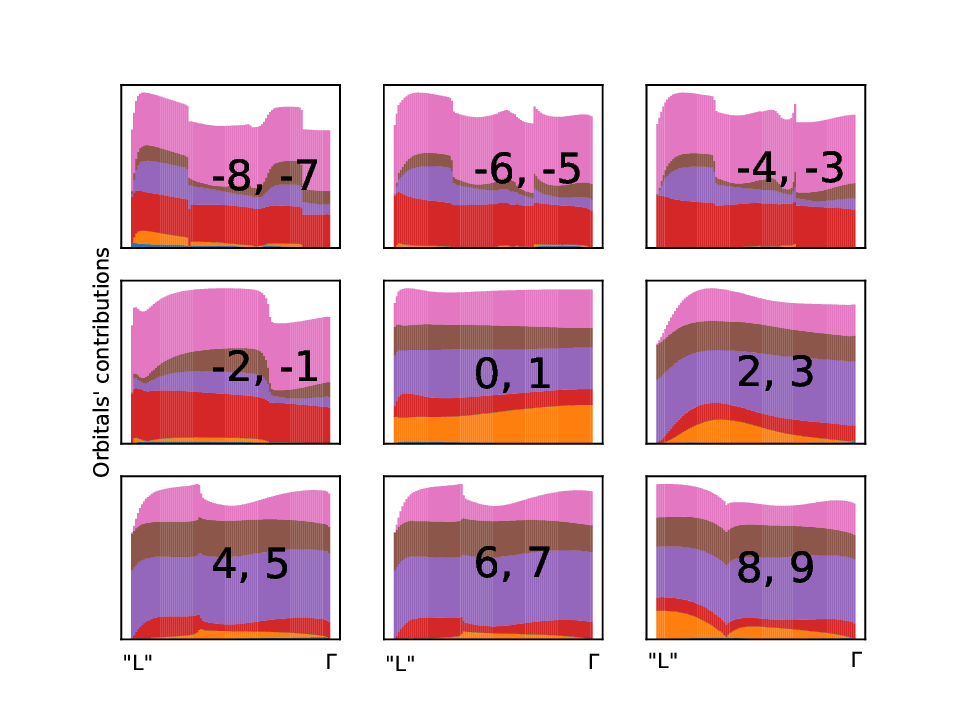}
  \caption{
Contributions of atomic orbitals to the wave functions in dependence on $k$ 
vector along [111] direction for PbTe:Bi. The consecutive bands are numbered 
as follows: 0 and 1 are the Bi induced gap states, bands from -8 to -1 are 
valence bands, and from 2 to 9 are conduction bands. 
The used color code is as
    follows: blue=6$s$(Bi), orange=6$p$(Bi), red=sum of $s$(Pb),
    violet=sum of $p$(Pb), brown= sum of $s$(Te), magenta= sum of
    $p$(Te).
   Because of the inversion symmetry, all bands are double degenerate,
 and the orbital composition of the spin partners (e.g., -8 and -7, -6
 and -5, etc) are identical. }
  \label{fig7}
\end{figure}

To understand the impact of Bi in more detail, we observe that the 6$p$(Bi) 
orbitals, which contribute to the Bi-induced gap state, are of odd parity, and 
accordingly they couple to the odd parity $L_6^-$ host wave functions. The 
"quantum repulsion" between these levels is clearly seen in the vicinity of the 
"L" point, Fig. \ref{fig5}(a). The coupling leads to the splitting of the 4-fold 
degenerate $L_6^{-}$ CBM at "L" into the almost unperturbed triplet and the 
singlet. The singlet is shifted upward by about 0.3 eV, whereas the Bi-induced 
band is down-shifted by the same amount. As a result, the coupling induces a 
relatively wide Bi-induced impurity band in the band gap, which is seen as a 
peak in the density of states in Fig. \ref{fig5}(b). One can also observe that 
without this $p$(Bi)--$L_6^-$ coupling, Bi ion would be a resonant donor 
degenerate with the conduction band. The Bi band is 0.3 eV wide, it partially 
overlaps with the CBM, and therefore Bi acts as a donor in agreement with 
experiment.

To analyze the Bi-related effects  we also project the relevant wave  
functions onto atomic orbitals. The spin degenerate partners occur in pairs,  
which are grouped in parantheses in the following. In the notation of Fig.  
\ref{fig7}, the (0,1) gap states are dominated by 6$p$(Bi) orbitals,
as expected.
Like in Fig. \ref{fig4} the discontinuities in the orbital content of most of the wave
functions presented in Fig. \ref{fig7} occur whenever two bands of
different symmetry intersect  as a function of the wave vector. \\
However, a comparable amount of $p$(Bi) is also found in the lowest  
conduction bands (2,3), and in bands (8,9) (especially at the "L" point) located  
0.3 eV above the CBM. This demonstrates that a strong hybridization takes  
place between 6$p$(Bi) and the conduction bands, and that at the "L" point  
the bands (8,9) almost have a character of a Bi-induced resonances. This  
hybridizaton is well visible in Fig.  \ref{fig7}, which shows the contribution of  
Bi orbitals to the host bands. The DOS projected onto the Bi orbitals is  
displayed in Fig. \ref{fig5}(b). The main and relatively sharp peak occurs ~0.1  
eV above the CBM, which shows that $E_{Fermi}$ lies within the  
conduction band, and the conductivity is of $n$-type. Finally, the VBM states of 
the $L_6^+$ symmetry are practically unperturbed  by the presence of Bi. The 
valence $5s$(Bi) states are situated a few eV below  the VBM and thus are 
absent in Fig. \ref{fig5}(a), they contribute to bonds  between the dopant and 
its neighbors.

The results for Bi in SnTe are shown in Fig. \ref{fig6}. Their symmetry-based 
interpretation is analogous to that for PbTe:Bi. In this case, this is the VBM of 
the $L_6^-$ symmetry which couples to the $6p$(Bi). Consequently, in the 
vicinity of the "L" point the splitting of the VBM occurs, the bands (-7,-8) are 
shifted down in energy by 0.2 eV, while the (0,1) Bi-related bands are shifted 
upwards by the same amount. The final result is the same as in PbTe: the Bi-
induced band is relatively wide, and it partially overlaps with the CBM, see Fig. 
\ref{fig6}(b). The strong coupling of 6$p$(Bi) with the host bands is clearly 
reflected in the orbital composition of the wave functions. Indeed, not only 
the (0,1) bands, but also the host bands close to VBM and CBM with 
appropriate symmetry contain appreciable amounts of $p$(Bi), indicating 
their efficient hybridization. In particular, we find a contribution of 6$p$(Bi) to 
the split-off bands (-8,-9). One can also notice that, coming back to the case of 
(Pb,Sn)Te, the impact of a Pb ion on the SnTe bands is limited to the CBM, 
again in agreement with the symmetry allowed $6p$(Pb)--$L_6^-$ coupling. 

From the results above it follows that the electronic states of a relatively 
shallow Bi donor cannot be described within the "classical" effective mass 
picture holding for shallow impurities, in which the impurity wave function 
consists in an envelope modulating the wave function from the CBM (or the 
VBM in the case of acceptors). In contrast, the Bi-induced states are strongly 
hybridized with the host states. This impurity-host coupling is to a large 
degree determined by the symmetry of the host states from the valence and 
conduction band extrema, and is effective only for the $L_6^-$ bands. This 
non-shallow character of Bi is surprising, since Bi is the "nearest neighbor" of 
Pb in the Periodic Table, and thus the central cell correction is expected to 
play a minor role. Finally, we recall that in a many-valley semiconductors,
such as Ge, a donor state is 
formed below every of the 4 CBMs at $L$. The valey-orbit splitting of these four-
fold degenerate donor state results in the formation of a singlet and a higher 
in energy triplet. From this point of view, both PbTe and SnTe are similar to 
Ge.

\subsection{Cr and Ag}   
\begin{figure}   
\includegraphics[width=\linewidth]{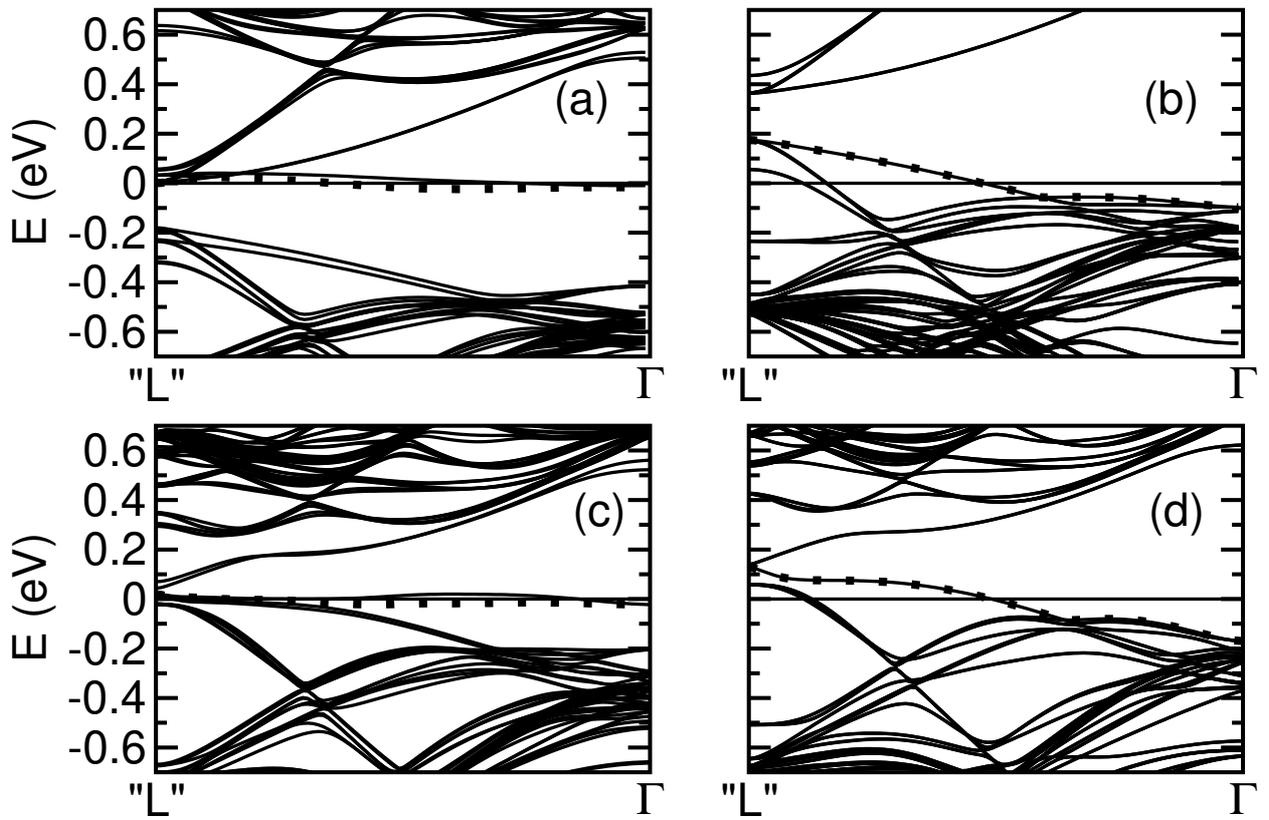}
   \caption{ 
Panels (a) and (b) show the bands of PbTe:Cr and PbTe:Ag, respectively. 
Panels (c) and (d) show the bands of  SnTe:Cr and SnTe:Ag, respectively. The 
dotted line show the bands, number of which is equal  to the number of 
electrons in the supercell. Energy zero is at $E_{Fermi}$.    
}   
\label{fig8} 
\end{figure}  

We now turn to the TM impurities, Cr and Ag. Before presenting and 
interpreting our results, we recall some features of TM atoms. First, the 
electronic configuration of an isolated Cr atom is $3d^54s^1$. In a rock salt 
host, the $d$-shell of the substitutional Cr is split by the octahedral crystal 
field into a $t_2$(Cr) triplet, and a $e_2$(Cr) doublet which is higher in energy 
by about 1--3 eV. Next, both multiplets are spin split by the exchange coupling. 
The Cr-derived spin-up bands are partially occupied and the spin polarization 
of Cr persists in both hosts, leading to a high spin state of Cr. In the presence 
of spin polarized Cr, the host bands are spin split.   

 The characteristics of a 4$d$ transition metal Ag is somewhat different. The 
electron configuration of Ag is $4d^{10}5s^1$ with the filled $d$-shell, which 
is a close analogue of Cu in the $3d^{10}4s^1$ configuration. One can notice 
that with the increasing atomic number of a TM dopant, the attractive nuclear 
Coulomb potential increases as well, which is reflected in the progressively 
lower energy of the $d$-shell electrons \cite{ciechan}. This effect is also 
present in our case: the energy of the $d$(Ag) shell is  lower than that of 
$d$(Cr), and both $t_2$(Ag) and $e_2$(Ag) are degenerate with the valence 
bands and completely occupied with electrons. The results for Cr and Ag  are 
summarized in one figure, Fig. \ref{fig8}, to simplify a comparison. In PbTe, 
the Cr-derived spin-up triplet is situated about 2 eV below the VBM, and is not 
shown in the Figure. As it is seen in Fig. \ref{fig8}(a), the $e_2$(Cr)-derived 
spin-up band overlaps with the bottom of the conduction band. This band is 
almost dispersionless, which reflects both the compactness of the 3$d$ 
orbitals and a weak hybridization with the conduction bands. Nominally, 
$e_2$(Cr) is half-filled, i.e., occupied with one electron. Consequently, Cr ions 
are in the Cr$^{2+}$ ($d^4$) states with the magnetic moment of about 4 
$\mu_B$. The 4$s$(Cr) states are well above the CBM.   

In the case of Ag in PbTe, Fig. \ref{fig8}(b), the $e_2$(Ag)-induced bands are 
situated 0.4 eV below the VBM, i.e., they are 0.6 eV lower in energy relative to 
those of Cr. $E_{Fermi}$ is about 0.18 eV below the VBM, which 
demonstrates that Ag acts as a resonant acceptor, and its $d$(Ag) shell is 
completely filled with 10 electrons. Consequently, the magnetic moment of 
Ag vanishes. This implies that there is no spin splitting of the host bands.

Turning to Cr and Ag in SnTe, Figs \ref{fig8}(c) and \ref{fig8}(d), respectively, 
we see that the energies of the dopant-derived bands are lower compared to 
those in PbTe. In particular, $e_2$(Cr) is close to the top of the valence bands, 
but neither the Cr$^{2+}$ charge state nor the magnetic moment are 
affected. Similarly, Ag is an efficient acceptor, and its magnetic moment 
vanishes.  Considering the impact of the dopants on the band gap we note 
that $E_{gap}$ of PbTe is unchanged to a very good approximation. In 
contrast, in SnTe from Figs \ref{fig8}(c) and \ref{fig8}(d) it follows that both Cr 
and Ag considerably modify the band gap. Indeed, its character changes from 
indirect to direct, and both VBM and CBM are at the "L" point. Importantly, 
the presence of either dopant reduces the band gap at "L" from 0.35 eV to 
0.05 eV. The effect is due to a strong modification of the CBM. Analysis of the 
conduction band wave functions reveals that, unlike in PbTe, the valence 
orbitals $s$(Cr) or $s$(Ag) largely contribute to the CBM, explaining its down 
shift. As it was discussed for Bi, the dopant-host hybridization effects are 
determined by symmetries of the involved states. In the case of Cr, both 
$s$(Cr) and $d$(Cr) orbitals are even with respect to inversion, and therefore 
they couple to even parity wave functions $L_6^+$. Accordingly, the 
symmetry based arguments explain the large, comparable to the band gaps, 
values of both the energy shifts of the $L_6^+$ band extrema and their 
splittings by ~0.2 eV. The same holds for Ag.

Our results compare well with these calculated based on the density functional theory 
reported in the literature. We use relatively large supercells with
216 atoms, while smaller 64- or 54-atom supercells used in some works
can distort the impurity-induced states, causing their broadening and
thus altering the doping features. However, the overall agreement is
satisfactory keeping in mind that the accuracy of ab initio methods is
also limited by the employed  exchange-correlation functionals, which
differ between teams. In particular, cation vacancies were found to be
relatively deep double acceptors  
degenerate with the valence bands in both PbTe and SnTe, Refs \cite{ahmad1, wang, bajaj, ryu, 
goyal, lee, hua, park, mishra}. These defects were extensively investigated 
because of their dominant impact on conductivity of both PbTe and
SnTe.

The same theoretical approach was used to assess properties of the dopants investigated 
here, and typically the results qualitatively agree with the present ones. This is the case of Ag 
found to be an acceptor \cite{ahmad, hoang, ryu}, which forms a
dispersive band close to the VBM in both hosts.
Experimentally, Ag indeed was found to be an efficient acceptor
\cite{osinniy}. Next,  according to \cite{hoang},  Bi in both PbTe and SnTe
induces a broad, 0.3 eV wide, donor band almost covering the entire
band gap, similar to our findings. The high doping efficiency of Bi is
confirmed in experiments with thermoelectric materials reporting Bi
n-type doping up to 10$^{19}$ - 10$^{20}$ cm$^{-3}$ \cite{dybko}. Finally, similar
to our finding, Cr is a resonant donor in PbTe  according to
Ref. \cite{xia}.

In the case of PbTe, our theoretical analysis is supported by experimental 
findings in the field of thermoelectricity. Indeed, a very high ($n,p$$\sim10^{19} 
- 10^{20}$ cm$^{-3}$) doping was successfully achieved: the $n$-type by 
doping with In, Bi and Cr on the cation sublattice, or by I doping on the anion 
sublattice. $p$-PbTe was obtained by Na, Tl or Ag acceptors substituting cations 
\cite{nimtz,khoklov,dybko,osinniy,paul,wojciechowski}.  

In undoped SnTe, $p$-type conductivity is observed with very high free hole 
concentrations ($p \sim 10^{20} - 10^{21}$ cm$^{-3}$). It originates from 
electrically active Sn vacancies, and dominates electrical and optical 
properties of SnTe. Any attempt to $n$-dope SnTe, e.g. with Bi,  Cr or Gd, 
resulted just in a reduction of hole concentration. Our analysis agrees well 
with these observation for Cr- or Bi-doped SnTe. 
Finally, in the case of SnTe:Ag we expect a large reduction of the band gap. 
This strong prediction requires experimental confirmation that, however, will 
be very challenging in view of exceptionally high native carrier concentration 
encountered in SnTe.

\section{Summary} 
\label{summary} 
We investigated the electronic properties of Bi, Ag, and Cr impurities and of  
cation vacancies in PbTe and SnTe by DFT calculations. Attention was focused 
on the symmetry-based interpretation of the results. As we show, it is the 
symmetry of the wave functions from the band extrema ($L_6^+$ or $L_6^-$) 
that has a decisive impact on the response of the host to a given 
dopant/defect. 

(i) Bi is a relatively shallow single donor in both PbTe and SnTe. However, it   
cannot be described within the effective mass picture holding for shallow  
impurities, because the Bi-induced states and the host states are strongly  
hybridized. This impurity-host coupling is symmetry-allowed only between  
$6p$(Bi) and $L_6^-$ bands (i.e., the CBM in PbTe, and the VBM in
SnTe).  The non-shallow character of Bi   
is surprising, since Bi is a neighbor to Pb in the Periodic Table, and  thus the 
central cell corrections are expected to play a minor role. Analogous   
symmetry arguments hold also for (Pb,Sn)Te. 

(ii) An analogous role of symmetry is found for Cr and Ag. In these cases, both  
$s$ and $d$ impurity orbitals are even with respect to inversion, and  
therefore they couple to the even parity wave functions $L_6^+$. Accordingly,  
the symmetry-based arguments explain the large, comparable to the band  
gaps, values of both the energy shifts of the $L_6^+$ band extrema and their  
splittings by ~0.2 eV. The doping properties of Cr and Ag are different: while  
Ag is an acceptor in both hosts, Cr is an acceptor in SnTe and a donor in PbTe. 

(iii) Finally, cation vacancies, abundant native defects in both hosts, are 
double acceptors. In the presence of the vacancy, the 4-fold degenerate VBM 
and CBM of both hosts are split, but both the sign and magnitude of the 
splittings depend on the symmetry of the corresponding wave function, and 
not on the host.
\ \\

{\bf ACKNOWLEDGMENTS.}
The work of  TS  was supported by the Foundation for Polish Science 
project "MagTop" no. FENG.02.01-IP.05-0028/23 co-financed by the European Union from the Funds 
of Priority 2 of the European Funds for a Smart Economy Program 2021-2027 (FENG) and by 
TechMatStrateg2/408569/5/NCBR/2019 project TERMOD of NCBR. The work of
A{\L} was partially supported by National Science
Centre NCN (Poland) project UMO-2017/27/B/ST3/02470.

\ \\

REFERENCES






\begin{thebibliography}{10}

\bibitem{nimtz}
G.~Nimtz, B.~Schlicht, Narrow-Gap Semiconductors, Springer-Verlag, Berlin,
  1983.

\bibitem{khoklov}
D.~R. Khokhlov (Ed.), Lead Chalcogenides: Physics and Applications, Taylor and
  Francis, New York, 2003.

\bibitem{hsieh}
T.~H. Hsieh, H.~Lin, J.~Liu, W.~Duan, A.~Bansil, L.~Fu, Nat. Commun. 3.

\bibitem{dziawa}
P.~Dziawa, B.~J. Kowalski, K.~Dybko, R.~Buczko, A.~Szczerbakow, M.~Szot,
  E.~{\L}usakowska, T.~Balasubramanian, B.~M. Wojek, M.~H. Berntsen,
  O.~Tjernberg, T.~Story, Nat. Mater. 11 (2012) 1023.

\bibitem{tanaka}
Y.~Tanaka, Z.~Ren, T.~Sato, K.~Nakayama, S.~Souma, T.~Takahashi, K.~Segawa,
  Y.~Ando, Nat. Phys. 8 (2012) 800.

\bibitem{xu}
S.~Y. Xu, C.~Liu, N.~Alidoust, M.~Neupane, D.~Quian, I.~Belopolski,
  J.~Denlinger, Y.~Wang, H.~Lin, L.~A. Wray, G.~Landolt, B.~Slomski, J.~H. Dil,
  A.~Marcinkova, E.~Morosan, Q.~Gibson, R.~Sankar, F.~C. Chou, R.~J. Cava,
  A.~Bansil, M.~Z. Hasan, Nat. Commun. 3 (2012) 1192.

\bibitem{dybko}
K.~Dybko, M.~Szot, A.~Mycielski, A.~Szczerbakow, M.~Guziewicz, W.~Knoff,
  E.~{\L}usakowska, T.~Story, Appl. Phys. Lett. 108 (2016) 133902.

\bibitem{osinniy}
V.~Osinniy, A.~Jedrzejczak, V.~Domukhovski, K.~Dybko, B.~Witkowska, T.~Story,
  Acta Phys. Pol. A 108 (2005) 809.

\bibitem{yu_cardona}
P.~Y. Yu, M.~Cardona, Fundamentals of Semiconductors, Springer Berlin
  Heidelberg New York, 2001.

\bibitem{Zunger}
S.-H. Wei, A.~Zunger, Phys. Rev. B 55 (1997) 13605.

\bibitem{hoang}
S.~K. Hoang, D.~Mahanti, M.~G. Kanatzidis, Phys. Rev. B 81 (2010) 115106.

\bibitem{ahmad}
S.~Ahmad, K.~Hoang, S.~D. Mahanti, Phys. Rev. Lett 96 (2006) 056403.

\bibitem{Tan}
X.~J. Tan, G.~Q. Liu, J.~T. Xu, H.~Z. Shao, J.~Jiang, H.~C. Jiang, Phys. Chem.
  Chem. Phys. 18 (2016) 20635.

\bibitem{Jovovic}
V.~Jovovic, S.~J. Thiagarajan, J.~P. Heremans, T.~Kommisarova, D.~Khokhlov,
  A.~Nicorici, J. Appl. Phys. 103 (2008) 053710.

\bibitem{Xiong}
K.~Xiong, G.~Lee, R.~P. Gupta, W.~Wang, B.~E. Gnade, K.~Cho, J. Phys. D: Appl.
  Phys. 43 (2010) 405403.

\bibitem{snyder}
G.~J. Snyder, E.~S. Toberer, Nat. Mater. 7 (2008) 105.

\bibitem{story92}
T.~Story, E.~Grodzicka, B.~Witkowska, J.~Gorecka, W.~Dobrowolski, Acta Phys.
  Polonica A 82 (1992) 879.

\bibitem{story93}
T.~Story, Z.~Wilamowski, E.~Grodzicka, B.~Witkowska, W.~Dobrowolski, Acta
  Physica Polonica A 84 (1993) 773.

\bibitem{grodzicka}
E.~Grodzicka, W.~Dobrowolski, J.~Kossut, T.~Story, B.~Witkowska, J. Cryst.
  Growth 138 (1994) 1034.

\bibitem{paul}
B.~Paul, P.~Rawat, P.~Banergij, Appl. Phys Lett. 98 (2011) 26211.

\bibitem{nielsen}
M.~D. Nielsen, E.~M. Levin, C.~M. Jaworski, K.~Schmidt-Rohr, J.~P. Heremans,
  Phys. Rev. B 85 (2012) 045210.

\bibitem{wang2}
F.~Wang, H.~Zhang, J.~Jiang, W.~L. Y.-F.~Zhao, J.~Yu, D.~Li, M.~H.~W. Chan,
  J.~Sun, Z.~Zhang, C.-Z. Chang, Phys. Rev. B 97 (2018) 115414.

\bibitem{skipetrov}
E.~P. Skipetrov, N.~A. Pichugin, B.~B. Kovalev, E.~I. Slyn'ko,
  V.~V.~E~Slyn’ko, Phys. B Condens. Matter 404 (2009) 5255.

\bibitem{wiendlocha}
J.~Heremans, B.~Wiendlocha, A.~M. Chamoire, Energy Environ. Sci. 5 (2012) 5510.

\bibitem{mahanti}
S.~D. Mahanti, K.~Hoang, S.~Ahmad, Physica B 401-402 (2007) 291.

\bibitem{ahmad1}
S.~Ahmad, S.~D. Mahanti, K.~Hoang, M.~G. Kanatzidis, Phys. Rev. B 74 (2006)
  155205.

\bibitem{CA}
D.~M. Ceperley, B.~J. Alder, Phys. Rev.Lett 45 (1980) 566.

\bibitem{openmx}
see http://www.openmx-square.org.

\bibitem{liu}
J.~Liu, T.~H. Hsieh, P.~Wei, W.~Duan, J.~Moodera, L.~Fu, Nat. Mat. 13 (2014)
  178.

\bibitem{ciechan}
A.~Ciechan, P.~Boguslawski, Optical Materials 79 (2018) 264.

\bibitem{wang}
N.~Wang, D.~West, J.~Liu, J.~Li, Q.~Yan, B.~L. Gu, S.~B. Zhang, W.~Duan, Phys.
  Rev. B 89 (2014) 045142.

\bibitem{bajaj}
S.~Bajaj, G.~S. Pomrehn, J.~W. Doak, W.~Gierlotka, H.~Wu, S.~W. Chen,
  C.~Wolverton, W.~A.~G. III, G.~J. Snyder, Acta Materialia 92 (2015) 72.

\bibitem{ryu}
B.~Ryu, M.-W. Oh, J.~K. Lee, J.~E. Lee, S.-J. Joo, B.-S. Kim, B.-K. Min,
  S.-D.~P. H.-W.~Lee, J. Appl. Phys. 118 (2015) 015705.

\bibitem{goyal}
A.~Goyal, P.~Gorai, E.~S. Toberer, V.~Stevanovi\'c, NPJ Computational Materials
  3 (2017) 42.

\bibitem{lee}
M.~H. Lee, S.~Park, J.~K. Lee, J.~Chung, B.~Ryu, S.-D. Parkc, J.-S. Rhyee, J.
  Matter. Chem. A 7 (2019) 16488.

\bibitem{hua}
F.~Hua, P.~Lv, M.~Hong, S.~Xie, M.~Zhang, C.~Zhang, W.~Wang, Z.~Wang, Y.~Liu,
  Y.~Yan, S.~Yuan, W.~Liu, X.~Tang, ACS Appl. Mater. Interfaces 13 (2021)
  56446.

\bibitem{park}
S.~Park, B.~Ryu, S.-D. Park, Materials 15 (2022) 1272.

\bibitem{mishra}
N.~Mishra, G.~Makov, J. Alloys Compd. 986 (2024) 174157.

\bibitem{xia}
M.~Xia, P.~Boulet, M.-C. Record, Phys. Rev. B 111 (2025) 045149.

\bibitem{wojciechowski}
K.~Wojciechowski, T.~Parashchuk, B.~Wiendlocha, O.~Cherniushok, Z.~Dashevsky,
  J. Mater. Chem. C 8 (2020) 13270.

\end{thebibliography}







\end{document}